\documentclass{emulateapj}
\usepackage{amsmath}
\usepackage{graphicx}
\usepackage{rotate}
\usepackage{graphics}
\usepackage{epsf,psfig}
\usepackage[dvips]{color}
\usepackage{color}
\usepackage{epsfig}
\usepackage{amssymb}
\usepackage{latexsym}
\usepackage{txfonts}

\newcommand{\un}[1]{~\hspace{-1pt}\ensuremath{\mathrm{#1}}}
\newcommand{\ti}[1]{$^{44}{}${#1}}
\newcommand{\ca}[1]{$^{40}{}${#1}}
\newcommand{\fe}[1]{$^{56}{}${#1}}

\newcommand{\integ}{{\it INTEGRAL}~}
\newcommand{\rxte}{{\it RXTE}~}
\newcommand{\ibis}{IBIS~}
\newcommand{\spi}{SPI~}
\newcommand{\isgri}{ISGRI~}
\newcommand{\ibisgri}{IBIS/ISGRI~}

\newcommand{\gro}{{\it CGRO}}
\newcommand{\sax}{{\it BeppoSAX}}

\newcommand{\gammaray}{$\gamma$-ray~}
\newcommand{\gammarays}{$\gamma$-rays~}
\newcommand{\xray}{X-ray~}

\def\d{$^\circ$}
\def\m{$^\prime$}
\def\s{$^{\prime\prime}$}
\def\hh{$^{\mathrm h}$}
\def\mm{$^{\mathrm m}$}
\def\ss{$^{\mathrm s}$}
\def\cm3{cm$^{-3}$}

\def\eg{{\it e.g.~}}
\def\etal{et~al.~}

\begin{document}

\title{The signature of \ti{Ti} in Cassiopeia A \\
       revealed by IBIS/ISGRI on INTEGRAL}

\slugcomment{Accepted version for ApJL on 29 June 2006}
\shorttitle{\ibisgri observations of Cassiopeia~A}
\shortauthors{Renaud \etal}

\author{
    M. Renaud\altaffilmark{1,2}, J. Vink\altaffilmark{3}, A. Decourchelle\altaffilmark{1,4},
    F. Lebrun\altaffilmark{1,2}, P.R. den Hartog\altaffilmark{5}, R. Terrier\altaffilmark{2},
    C. Couvreur\altaffilmark{1}, J. Kn\"{o}dlseder\altaffilmark{6}, P. Martin\altaffilmark{6},
    N. Prantzos\altaffilmark{7}, A.M. Bykov\altaffilmark{8}, H. Bloemen\altaffilmark{5}
    }

\altaffiltext{1}{\scriptsize Service d'Astrophysique,
DAPNIA/DSM/CEA, 91191 Gif-sur-Yvette, France; mrenaud@cea.fr}

\altaffiltext{2}{\scriptsize APC-UMR 7164, 11 place M. Berthelot,
75231 Paris, France}

\altaffiltext{3}{\scriptsize Astronomical Institute, Utrecht University,
P.O. Box 80000, 3508 TA Utrecht, The Netherlands}

\altaffiltext{4}{\scriptsize AIM-UMR 7158, 91191 Gif-sur-Yvette, France}

\altaffiltext{5}{\scriptsize SRON Netherlands Institute for Space
Research, Sorbonnelaan 2, 3584 AC Utrecht, The Netherlands}

\altaffiltext{6}{\scriptsize Centre d'Etude Spatiale des
Rayonnements and Universit\'e Paul Sabatier, 31028 Toulouse,
France}

\altaffiltext{7}{\scriptsize Institut d'Astrophysique de Paris,
75014 Paris, France}

\altaffiltext{8}{\scriptsize A.F. Ioffe Institute for Physics and
Technology, St. Petersburg, Russia, 194021}

\begin{abstract}

We report the detection of both the 67.9 and 78.4\un{keV} \ti{Sc}
\gammaray lines in Cassiopeia~A with the \integ \ibisgri
instrument. Besides the robustness provided by spectro-imaging
observations, the main improvements compared to previous
measurements are a clear separation of the two \ti{Sc} lines
together with an improved significance of the detection of the
hard \xray continuum up to 100\un{keV}. These allow us to refine
the determination of the \ti{Ti} yield and to constrain the nature
of the nonthermal continuum emission. By combining COMPTEL,
\sax/PDS and \isgri measurements, we find a line flux of (2.5
$\pm$ 0.3) $\times$ 10$^{-5}$ cm$^{-2}$ s$^{-1}$ leading to a
synthesized \ti{Ti} mass of 1.6 $^{+0.6}_{-0.3}$ $\times$
10$^{-4}$ M$_{\odot}$. This high value suggests that Cas~A is
peculiar in comparison to other young supernova remnants, from
which so far no line emission from \ti{Ti} decay has been
unambiguously detected.

\end{abstract}


%_________________________________________________________________________________________________

\section{Introduction}
\label{s:intro}

Cassiopeia~A (hereafter, Cas~A) is the youngest known supernova
remnant (SNR) in the Milky Way, located at a distance of
3.4$^{+0.3}_{-0.1}$\un{kpc} \cite{c:reed95}. The estimate of the
supernova is A.D.~1671.3$\pm$0.9, based on the proper motion of
several ejecta knots \cite{c:thorstensen01}. However, an event
observed by Flamsteed (A.D.~1680) could be at the origin of the
Cas~A remnant \cite{c:ashworth80,c:stephenson02}. The large
collection of data from observations in the radio, infra-red,
optical, \xray (see \eg Hwang \etal 2004) up to TeV \gammarays
\cite{c:aharonian01} allows us to study its morphology,
composition, cosmic-ray acceleration efficiency and secular
evolution in details. Young SNRs are thought to be efficient
particle accelerators and represent the main galactic production
sites of heavy nuclei, some of them being radioactives. Soft
\gammaray observations, beyond the thermal \xray emission ($\geq$
10\un{keV}), can therefore provide invaluable information in both
of these areas by studying the nonthermal continuum and the
\gammaray line emission. Cas~A then appears to be the best case
for such investigations.

Few radioactive isotopes are accessible to \gammaray astronomy for
probing cosmic nucleosynthesis \cite{c:diehl98}. Amongst them,
\ti{Ti} is a key isotope for the investigation of the inner
regions of core-collapse SNe and their young remnants. This
nucleus is thought to be exclusively created in SNe but with a
large variation of yields depending on their type. Recent accurate
measurements by several independent groups give a weighted-average
\ti{Ti} lifetime of 86.0 $\pm$ 0.5 years
\cite{c:ahmad98,c:gorres98,c:norman98,c:wietfeldt99,c:hashimoto01}.
The discovery of the 1157\un{keV} \ti{Ca} $\gamma$-ray line
emission from the decay chain of \ti{Ti}
($^{44}$Ti$\longrightarrow$$^{44}$Sc$\longrightarrow$$^{44}$Ca)
with \gro/COMPTEL \cite{c:iyudin94} was the first direct proof
that this short-lived isotope is indeed produced in SNe. This has
been strengthened by the \sax/PDS detection of the two blended low
energy \ti{Sc} lines at 67.9\un{keV} and 78.4\un{keV}
\cite{c:vink01}. By combining both observations, Vink \etal (2001)
deduced a \ti{Ti} yield of (1.5$\pm$1.0) $\times$ 10$^{-4}$
M$_{\odot}$.

This high value compared to those predicted by "standard" models
(\eg Woosley \& Weaver 1995b, WW95; Thielemann, Nomoto, \& Hashimoto 1996,
TNH96) as well as improved ones \cite{c:rauscher02,c:limongi03} could be
due to several effects. First of all, the explosion of Cas~A seems to
have been intrinsically asymmetric since such asymmetries have recently
been observed in the ejecta \cite{c:vink04,c:hwang04}, and there are
indications that its explosion energy was $\sim$ 2 $\times$ 10$^{51}$
erg \cite{c:laming03}, higher than the canonical value of 10$^{51}$
erg. The sensitivity of the \ti{Ti} production to the explosion
energy and asymmetries may explain the high \ti{Ti} yield compared
to explosion models \cite{c:nagataki98}.

It is generally accepted that Cas~A was formed by the explosion of a
massive progenitor, from a 16 M$_{\odot}$ single star \cite{c:chevalier03}
to a Wolf-Rayet (WR) remnant of a very massive ($<$ 60 M$_{\odot}$)
precursor \cite{c:fesen91}. Type Ib explosions, originating from
progenitors which have experienced strong mass loss (see Vink
2004, Vink 2005), should on average produce more \ti{Ti} due to
the lower fall back of material on the compact stellar remnant
\cite{c:wlw95a}. However, there is some debate on the detailed
stellar evolution scenario that may have accounted for the low
mass of the star prior to the explosion. The amount of oxygen
present (1-2 M$_{\odot}$, Vink \etal 1996) suggests a main
sequence mass of 20 M$_{\odot}$. This may be too low to form a
Type Ib progenitor by mass loss in a WR phase. Moreover, the high
surrounding density is better explained if the shock wave is
moving through the dense wind of a red supergiant rather than the more
tenuous wind of a WR. Therefore, it has been recently suggested that the
low mass of the progenitor is the result of a common envelope evolutionary
phase in a binary system \cite{c:young06}. The authors demonstrated that
such a scenario of a 15-25 M$_{\odot}$ progenitor which lost its
hydrogen envelope due to a binary interaction can match the main
observational constraints. In any case, the \ti{Ti} production is
highly sensitive to details of the explosion as well as nuclear
reaction rates. It is of interest to point out that the major
\ti{Ti} production reaction \ca{Ca}($\alpha$,$\gamma$)\ti{Ti} has
been revised \cite{c:nassar06}, implying an increase of the
\ti{Ti} production by a factor of $\sim$ 2.

In addition to the \ti{Sc} \gammaray lines, the hard \xray
spectrum is also of interest for its nonthermal continuum emission
and because this underlying continuum is critical to properly
measure the \ti{Sc} line flux. Nevertheless, its nature is still
under debate. The nonthermal hard \xray continuum could be due to
either synchrotron radiation of TeV electrons \cite{c:allen97} or
nonthermal bremsstrahlung from supra-thermal electrons which have
been accelerated by internal shocks (Laming 2001a,b; Vink \&
Laming 2003). Both cases predict a gradual steepening at high
energies and then, reliable continuum flux measurements beyond the
two low energy \ti{Sc} lines ($>$ 80\un{keV}) are necessary, as
initiated with \gro/OSSE \cite{c:the96}. Soft \gammaray
observations are therefore critical to better understand the
nucleosynthesis and the particle acceleration processes in young
SNRs such as Cas~A. \ibis \cite{c:ubertini03}, one of the two main
coded mask aperture instruments onboard the \integ satellite
\cite{c:winkler03}, is best suited to study both the hard \xray
continuum and the line emission thanks to its low energy
(15\un{keV} -- 1\un{MeV}) camera \isgri \cite{c:lebrun03}.
\ibisgri provides spectro-imaging (13\m~FWHM, 6\un{keV} FWHM at
70\un{keV}) over a large field of view (400 deg$^{2}$) in the
energy range 15\un{keV}-1\un{MeV} with a milliCrab sensitivity at
70 keV (3 $\sigma$, $\Delta$E/E = 2, 10$^{6}$ s). The large field
of view allows for long exposures devoted to the simultaneous
observation of several sources. In this letter, we report the
results of the spectro-imaging analysis of Cas~A based on \ibisgri
observations.

%_________________________________________________________________________________________________

\section{INTEGRAL/IBIS observations and Data Analysis}
\label{s:obs}

Since its launch, \integ has performed deep Open Time observations
dedicated to the Cassiopeia region, mainly for measuring and
constraining the \ti{Ti} production in the Cas~A and Tycho SNRs.
Preliminary results on these two young SNRs are reported in Vink
\etal (2005) and Renaud \etal (2006), respectively. Moreover, den
Hartog \etal (2006) have presented a comprehensive list of the
sources detected by \ibisgri above 20\un{keV} in this region. We
have performed a detailed analysis of $\sim$ 1800 pointings or
science windows (hereafter, scws), each of them lasting typically
between 1800 and 3500 s during which the telescopes are pointed at
a fixed direction. We have selected pointings at less than
11\d~from Cas~A and removed those for which the Veto and \isgri
($\geq$ 500\un{keV}) counting rates were above 3.5 $\times$
10$^{4}$ and 45 ct.s$^{-1}$, respectively. The total effective
time is then $\sim$ 3.2\un{Ms} (over $\sim$ 4.5\un{Ms} of total
exposure time).

For \gammaray line studies, the most critical part of the
\ibisgri data analysis is the energy correction of detected
events. The spectral performance of the \isgri camera depends on
the alignment of the pixel gains and offsets. Based on more than
two years of observations, a fine in-flight calibration has been
done by taking into account several parameters such as the
temperature, the accumulated proton irradiation, and the time
after the detector switch-on. Moreover, because of the charge loss
in the Cadmium Telluride (CdTe) detectors and in their electronics,
the \isgri spectral response above $\sim$ 60\un{keV} depends on the pulse
rise-time and a second software correction is needed \cite{c:lebrun03}.

\begin{figure}

\includegraphics[scale=0.36, angle=0]{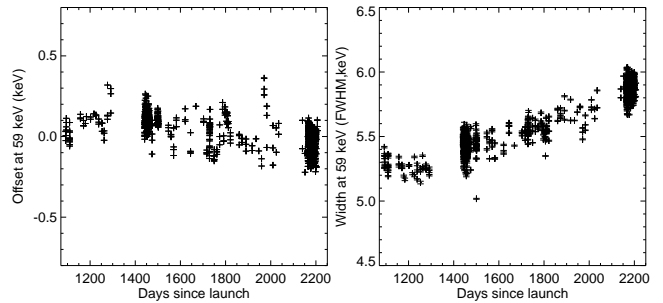}
\caption{Position (left) and width (right) of the W K$_{\alpha}$
background line.\label{f:isgri_lines}}

\end{figure}

\begin{figure*}

\includegraphics[scale=0.7, angle=0]{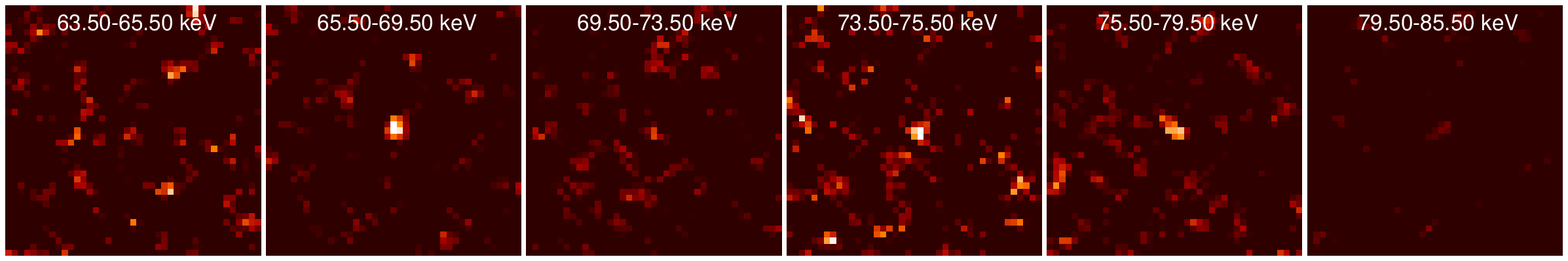}
\caption{\ibisgri flux images centered on Cas~A (2.5\d $\times$
2.5\d) in six energy bands. The linear scale is the same for all
images, between 10$^{-6}$ and 4.7 $\times$ 10$^{-6}$ cm$^{-2}$
s$^{-1}$ keV$^{-1}$. Note that the noise in the images depends on
the energy band widths. \label{f:isgri_ima}}

\end{figure*}

To evaluate the efficiency of all these corrections, we measured
the position and the width of the W K$_{\alpha}$ fluorescence
background line at 59\un{keV} for each scw. As shown in Figure
\ref{f:isgri_lines}, the dispersion of the 59\un{keV} line
position over the 3 years of observations is about 0.1\un{keV}.
The spectral degradation observed on the right panel of Figure
\ref{f:isgri_lines} is due to the irradiation of the detector
pixels but is still negligible after 3 years in terms of line
sensitivity ($\sim$ 5\%). The deconvolution of coded mask images
(shadowgrams) removes completely the background only if it is
flat. Background structures in the shadowgram produces large scale
structures in the deconvolved image. To avoid them, a background
map is first subtracted from the shadowgram. Such correcting
background maps were produced by summing a large number of high
latitude observations from all directions. In this way, the
shadowgrams of the many weak sources are smeared out on the
detector. With more than 2\un{Ms} of exposure time, these ensure
the best removal of structures in the detector images, mainly
around the fluorescence lines located close to the two low energy
\ti{Sc} astrophysical lines. We then used the Off-Line Scientific
Analysis (OSA) software \cite{c:goldwurm03} version 5.1 in order
to obtain sky images and we have constructed mosaic images in 14
energy bands (see Figs. \ref{f:isgri_ima} and \ref{f:isgri_spec}).

%_________________________________________________________________________________________________

\section{Results}
\label{s:res}

In order to estimate the source position of the hard \xray
continuum, we also analyzed the 18-25\un{keV} energy band which
has the best signal to noise ratio (hereafter, S/N) for a steep
spectrum such as that of Cas~A. We have fitted the source with a
two-dimensional elliptical gaussian with the following parameters:
the background level, the position and the value of the maximum,
the widths on the two axes, and the rotation angle of the ellipse.
We did not find any evidence of a source extent (the two widths
are close to 14\m~FWHM). The fitted position of Cas~A is R.A. =
23\hh~23\mm~22.6\ss, decl. = +58\d~49\m~02.1\s~(J2000) with a S/N
of $\sim$ 38. According to Gros \etal (2003), the corresponding
point source location error (PSLE) radius at the 90\% confidence
level is $\sim$ 50\s. Therefore, the full error box is contained
within the remnant.

\begin{figure}[htb]

\includegraphics[scale=0.355, angle=0]{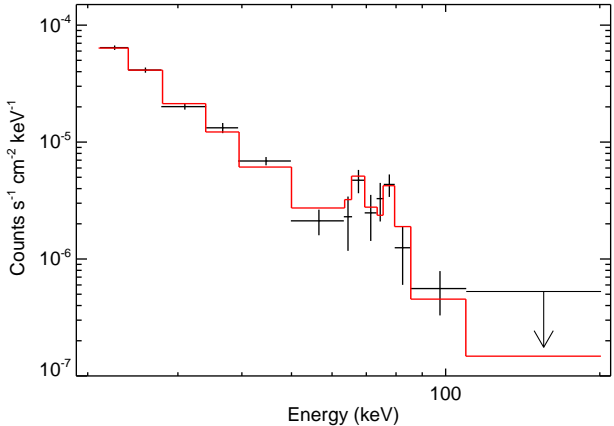}
\caption{\ibisgri spectrum of Cas~A and the best-fit model as
described in the text (solid red line) with the following
boundaries: 21, 24, 28, 34, 39.5, 50, 63.5, 65.5, 69.5, 73.5,
75.5, 79.5, 85.5, 109.5 and 201\un{keV}. The count rates and the
model have been divided by the effective \isgri area at the center
of each channel, in order to obtain approximate flux density
units. The upper limit above 110\un{keV} is given at the 3
$\sigma$ confidence level. \label{f:isgri_spec}}

\end{figure}

Figure \ref{f:isgri_ima} shows \ibisgri images centered on Cas~A
in the six energy bands around the two \ti{Sc} lines that shows
that the source brightens at the line energies. For building up
the source spectrum, we first measured in each individual sky
image the flux and its associated variance at the pixel
corresponding to the fitted position in the 18-25\un{keV} energy
range. Note that this variance takes into account all
uncertainties, in particular those resulting from the background
subtraction. We then calculated the weighted mean count rate and
corresponding error for each of the 14 energy bands. This spectrum
is presented in Figure \ref{f:isgri_spec}, showing the clear
detection of the two low energy \ti{Sc} lines. We tested two
different models for the continuum emission: the pegged power-law
{\tt pegpwrlw} in the 21-120\un{keV} band and the {\tt srcut}
\cite{c:reynolds99} model in XSPEC v.11.3. This latter is an
approximation of the \xray synchrotron radiation from young SNRs.
The \ti{Sc} lines were fitted with two gaussians of equal
intensity at fixed positions and with no line broadening.

\begin{deluxetable*}{cccccccc}
\tabletypesize{\scriptsize} \tablecaption{Spectral model fits}
\tablehead{
\colhead{} & \colhead{} & \colhead{} & \colhead{Total flux in the} & \colhead{} & \colhead{} & \colhead{Roll-off} & \colhead{} \\
\colhead{} & \colhead{\ti{Sc} Flux} & \colhead{Power-Law} & \colhead{21-120\un{keV} range} & \colhead{Flux density} & \colhead{} & \colhead{Energy} & \colhead{} \\
\colhead{Model} & \colhead{(10$^{-5}$ ph cm$^{-2}$ s$^{-1}$)} &
\colhead{Index} & \colhead{(10$^{-12}$ erg cm$^{-2}$ s$^{-1}$)} &
\colhead{at 1\un{GHz} (Jy)} & \colhead{Radio Index} &
\colhead{(keV)} & \colhead{$\chi^{2}$/$\nu$} } \startdata
Power-Law & 2.2 $\pm$ 0.5 &  3.3 $\pm$ 0.1  & 37.5 $\pm$ 1.5  &      ---        &      ---        &       ---       &  9.5/10  \\
SRCUT     & 2.9 $\pm$ 0.5 &      ---        &       ---       &
2720 (fixed)  &   0.77 (fixed)  & 0.97 $\pm$ 0.02 & 18.5/11
\enddata
\label{t:res_sp}

\end{deluxetable*}

\begin{figure}

\includegraphics[scale=0.185]{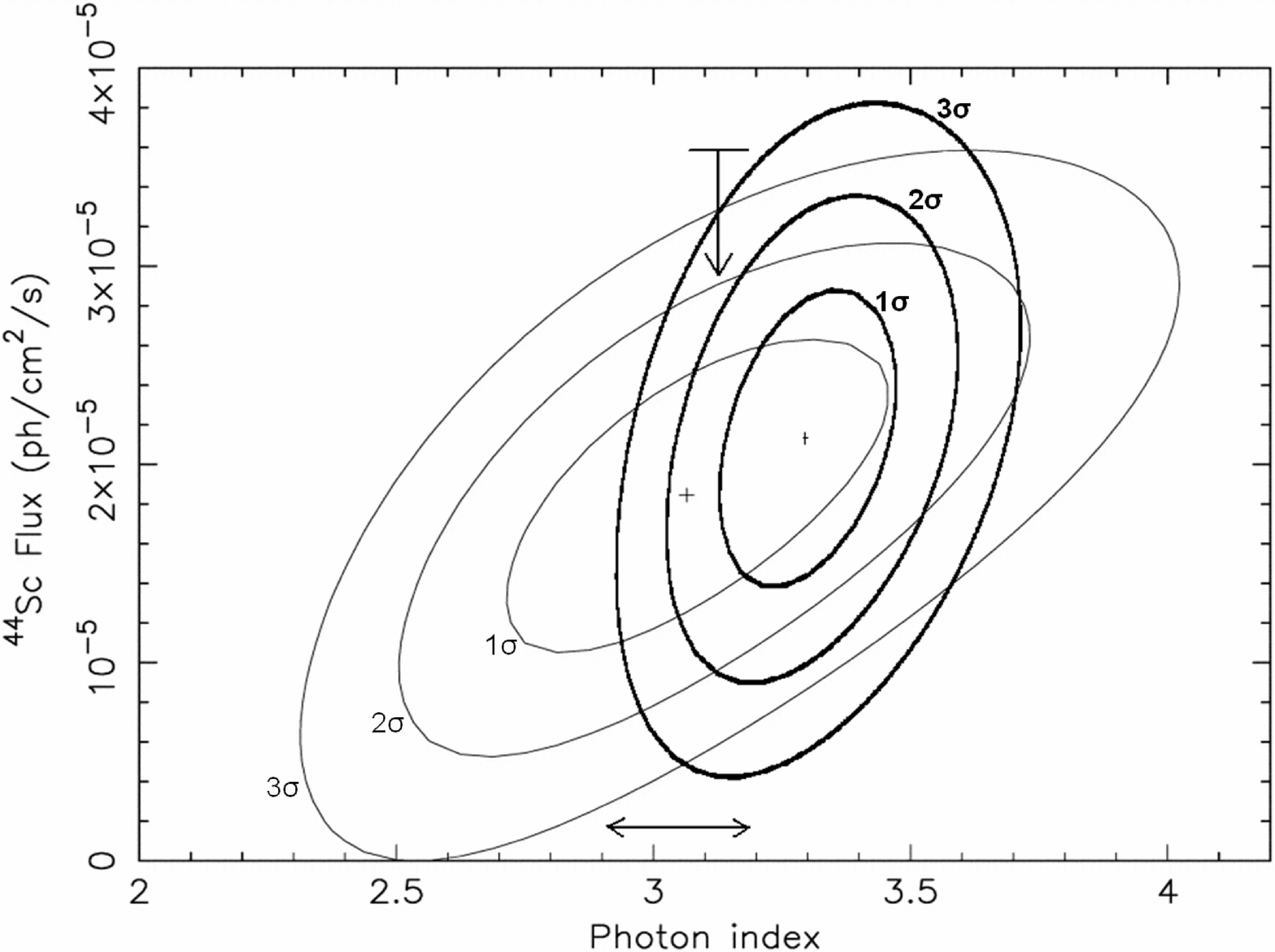}
\caption{Confidence ellipses for the combination of the \ti{Sc}
line flux and the power-law photon index in the 21-120\un{keV}
band with \ibisgri (thick lines), and in the 30-100\un{keV} band
with \sax/PDS (thin lines, Vink \etal 2001). The vertical arrow
corresponds to the \rxte upper limit at 90 \% confidence level on
the \ti{Sc} line flux \cite{c:rothschild03} and the horizontal
arrow presents the first \rxte and OSSE measurements of the photon
index \cite{c:allen97}. \label{f:correl_ima}}

\end{figure}

The obtained best-fit parameters together with their 1 $\sigma$
confidence levels are given in Table \ref{t:res_sp}. The best-fit
model is found with a pure power-law continuum spectrum and a
\ti{Sc} line flux of (2.2 $\pm$ 0.5) $\times$ 10$^{-5}$ cm$^{-2}$
s$^{-1}$ in each line ($\chi^{2}$ = 9.5 for 10 dof) consistent
with previous COMPTEL and \sax/PDS measurements. Taken together
the \ti{Sc} lines are detected at the 4.5 $\sigma$ confidence
level ($\Delta$$\chi^{2}$ = 20) and each is individually detected
at 3 $\sigma$ above the continuum emission. Relaxing the
constraints on the line positions and width results in fitted
line flux (2.3 $\pm$ 0.5 $\times$ 10$^{-5}$ cm$^{-2}$ s$^{-1}$), positions
(67.8 $\pm$ 1.6 and 77.4 $\pm$ 1.4\un{keV}) and width ($<$
1.6\un{keV}) consistent with the expected values and does not
improve the fit ($\chi^{2}$ = 9.5 for 7 dof). The 3 $\sigma$ upper
limit on the line broadening translates to a non-constraining
upper limit of $\sim$ 14,000 km s$^{-1}$ for the expansion velocity.
For the {\tt srcut} model, the flux density at 1\un{GHz} and the
radio spectral index were fixed to 2720\un{Jy} and 0.77
\cite{c:green05}. We find a $\chi^{2}$ of 18.5 for 11 dof.
According to the F-test, a power-law is favored over the {\tt
srcut} model at 2.5 $\sigma$ (98.8 \%). It would be also the case
of any other model that predicts a substantial steepening of the
continuum emission above 50\un{keV}. From Table \ref{t:res_sp} it
is clear that the estimate of the \ti{Sc} line flux is sensitive
to this continuum modeling and we then explored the correlation
between the \ti{Sc} line flux and the power-law photon index.
Figure \ref{f:correl_ima} presents such a correlation diagram. A 
detailed analysis of the nature of the hard \xray continuum, its
effect on the \ti{Sc} line flux estimate, and the results obtained
with the {\it INTEGRAL}/\spi data will be presented in a forthcoming
paper (Vink \etal 2006, in preparation).

%_________________________________________________________________________________________________

\section{Discussion}
\label{s:discuss}

The \ibisgri observations confirm the presence of the two low energy
\ti{Sc} \gammaray lines in Cas~A. By performing a weighted average of
the three independent measurements of COMPTEL, \sax/PDS \cite{c:vink01}
and \isgri, we find a line flux of (2.5 $\pm$ 0.3) $\times$ 10$^{-5}$
cm$^{-2}$ s$^{-1}$. Taking into account uncertainties on its age \cite{c:thorstensen01},
distance \cite{c:reed95} and \ti{Ti} lifetime \cite{c:vink05}, this is
translated into an initial synthesized \ti{Ti} mass of
(1.6$^{+0.6}_{-0.3}$) $\times$ 10$^{-4}$ M$_{\odot}$. This mass of
ejected \ti{Ti} is generally thought to be unusually large (or
for few specific cases, marginally consistent) in comparison with
spherical explosion models of WW95 and TNH96 \cite{c:timmes96}.
Moreover, in the standard frame where \ti{Ti} and \fe{Ni} are
co-produced during the first stages of the explosion, Cas~A should
have been a very bright, \fe{Ni}-rich SN, in contrast with its non
detection or with the Flamsteed's historical record. However, the large
\ti{Ti}/\fe{Ni} ratio could be explained by the high degree of
asymmetries \cite{c:nagataki98}.The high \ti{Ti} yield thus supports the
idea that Cas~A is the result of an asymmetric and/or a relatively more
energetic explosion, consistent with other observational evidence
\cite{c:vink04,c:hwang04}.

Anyway, the \ti{Ti} production in core-collapse SNe is highly sensitive to
the network used to compute nuclear reactions. With the recent revised
\ca{Ca}($\alpha$,$\gamma$)\ti{Ti} reaction rate \cite{c:nassar06},
theoretical models become more compatible with the \ti{Ti} yield
deduced from \ibisgri and previous observations. However, this
would make the lack of other Galactic \ti{Ti} sources an even more
serious problem: several \gammaray line surveys \cite{c:dupraz97,c:renaud04,c:the06}
have highlighted the problem of the "young, missing, and hidden" Galactic SNe,
those that should have occurred since Cas~A and are still not detected through the
line emission from \ti{Ti} decay. This would strengthen the idea that Cas~A
is peculiar \cite{c:young06}. On the other hand, the high \ti{Ti} yield
of both Cas~A and SN~1987A \cite{c:fransson01} is more in accordance with
the solar \ti{Ca}/\fe{Fe} ratio, whereas this ratio is underpredicted by
current spherically symmetric explosive nucleosynthesis models
\cite{c:prantzos04,c:young06}.

Besides the robustness provided by these \ibisgri spectro-imaging
observations, the main improvements compared to previous
observations \cite{c:vink01,c:rothschild03} are the improved
spectral resolution and the improved significance of the detection
of the hard \xray nonthermal continuum up to 100\un{keV} well
fitted by a single power-law. The latter gives more stringent
constraints on both the line intensities and the underlying
continuum. Therefore, the scenario of a synchrotron radiation by
TeV electrons \cite{c:allen97} as modeled by Reynolds \& Keohane
(1999) seems not appropriate in the case of Cas~A. On the other
hand, the model developed by Laming (2001a,b) implying a
nonthermal bremsstrahlung emission of supra-thermal electrons
could be an alternative scenario. Based on this firm detection of
the \ti{Sc} lines with IBIS/ISGRI, the expected results with SPI,
thanks to its fine spectral resolution ($\Delta$E $\sim$ 2\un{keV}
FWHM at 1\un{MeV}), should help us for the first time to constrain
the kinematics of the innermost layers of the explosion (Vink
\etal 2006, in preparation).

%_________________________________________________________________________________________________

\acknowledgements{The present work is based on observations with
{\it INTEGRAL}, an ESA project with instruments and science data
center (ISDC) funded by ESA members states (especially the PI
countries: Denmark, France, Germany, Italy, Switzerland, Spain,
Czech Republic and Poland, and with the participation of Russia
and the USA). \isgri has been realized and maintained in flight by
CEA-Saclay/DAPNIA with the support of CNES.}


\begin{thebibliography}{dummy}

\bibitem[Aharonian \etal 2001]{c:aharonian01} Aharonian, F., \etal 2001, \aap, 370, 112
\bibitem[Ahmad \etal 1998]{c:ahmad98} Ahmad, I., \etal 1998, \prl, 80, 2559
\bibitem[Allen \etal 1997]{c:allen97} Allen, G.~E., \etal 1997, \apjl , 487, L97
\bibitem[Ashworth 1980]{c:ashworth80} Ashworth, W.~B. 1980, J.~Hist.~Astron., 11, 1
\bibitem[Chevalier \& Oishi 2003]{c:chevalier03} Chevalier, R.~A., \& Oishi, J. 2003, \apj, 593, 23
\bibitem[Den Hartog \etal 2006]{c:denhartog06} den Hartog, P.~R., Hermsen, W., Kuiper, L., Vink, J., in 't Zand, J.~J.~M., \& Collmar, W. 2006, \aap, 451, 587
%\bibitem[Den Hartog \etal 2006]{c:denhartog06} den Hartog, P.~R., \etal 2006, \aap, 451, 587
\bibitem[Diehl \& Timmes 1998]{c:diehl98} Diehl, R., \& Timmes, F.~X. 1998, \pasp, 110, 637
\bibitem[Dupraz \etal 1997]{c:dupraz97} Dupraz, C., \etal 1997, \aap, 324, 683
\bibitem[Fransson \& Kozma 2001]{c:fransson01} Fransson, C., \& Kozma, C., 2001, New Astr. Rev., 46, 487
\bibitem[Fesen \& Becker 1991]{c:fesen91} Fesen, R.~A., \& Becker, R.~H. 1991, \apj, 371, 621
\bibitem[Goldwurm \etal 2003]{c:goldwurm03} Goldwurm, A., \etal 2003, \aap, 411, L223
\bibitem[G\"{o}rres \etal 1998]{c:gorres98} G\"{o}rres, J., \etal 1998, \prl, 80, 2554
\bibitem[Green 2005]{c:green05} Green, D.A. 2005, astro-ph/0505428
\bibitem[Gros \etal 2003]{c:gros03} Gros, A., \etal 2003, \aap, 411, L179
\bibitem[Hashimoto \etal 2001]{c:hashimoto01} Hashimoto, T., \etal 2001, Nucl.~Phys.~A, 686, 591
\bibitem[Hwang \etal 2004]{c:hwang04} Hwang, U., \etal 2004, \apj, 615, L117
\bibitem[Iyudin \etal 1994]{c:iyudin94} Iyudin, A.~F., \etal 1994, \aap, 284, L1
\bibitem[Laming 2001a]{c:laming01a} Laming, J.~M. 2001a, \apj, 546, 1149
\bibitem[Laming 2001b]{c:laming01b} Laming, J.~M. 2001b, \apj, 563, 828
\bibitem[Laming \& Hwang 2003]{c:laming03} Laming, J.~M., \& Hwang, U. 2003, \apj, 597, 347
\bibitem[Lebrun \etal 2003]{c:lebrun03} Lebrun, F., \etal 2003, \aap, 411, L141
\bibitem[Limongi \& Chieffi 2003]{c:limongi03} Limongi, M., \& Chieffi, A. 2003, \apj, 592, 404
\bibitem[Nagataki \etal 1998]{c:nagataki98} Nagataki, S., Hashimoto, M., Sato, K., Yamada, S., \& Mochizuki, Y. 1998, \apj, 492, 45
\bibitem[Nassar \etal 2006]{c:nassar06} Nassar, H., \etal 2006, \prl, Volume 96, Issue 4
\bibitem[Norman \etal 1998]{c:norman98} Norman, E.~B., \etal 1998, \prc, 57, 2010
\bibitem[Prantzos 2004]{c:prantzos04} Prantzos, N. 2004, 5th INTEGRAL Workshop on the INTEGRAL Universe (ESA SP-552), 16-20 February 2004, Munich, Germany
%\bibitem[Rauscher \etal 2002]{c:rauscher02} Rauscher, T., Heger, A., Hoffman, R.~D., \& Woosley, S.~E. 2002, \apj, 576, 323
\bibitem[Rauscher \etal 2002]{c:rauscher02} Rauscher, T., \etal 2002, \apj, 576, 323
\bibitem[Reed \etal 1995]{c:reed95} Reed, J.~E., Hester, J.~J., Fabian, A.~C., \& Winkler, P.~F. 1995, \apj, 440, 706
\bibitem[Renaud \etal 2004]{c:renaud04} Renaud, M., Lebrun, F., Ballet, J., Decourchelle, A., Terrier, R., \& Prantzos, N. 2004, 5th INTEGRAL Workshop on the INTEGRAL Universe (ESA SP-552), 16-20 February 2004, Munich, Germany
\bibitem[Renaud \etal 2006]{c:renaud06} Renaud, M., Vink, J., Decourchelle, A., Lebrun F., Terrier, R., \& Ballet, J. 2006, AwR V conference, 5-9 September 2005, Clemson, USA
\bibitem[Reynolds \& Keohane 1999]{c:reynolds99} Reynolds, S.~P., \& Keohane, J.~W. 1999, \apj, 525, 368
\bibitem[Rothschild \& Lingenfelter 2003]{c:rothschild03} Rothschild, R.~E., \& Lingenfelter, R.~E. 2003, \apj, 582, 257
\bibitem[Stephenson \& Green 2002]{c:stephenson02} Stephenson, F.~R., \& Green, D.~A. 2002, Historical Supernovae and their Remnants (Oxford: OUP)
\bibitem[Timmes \etal 1996]{c:timmes96} Timmes, F.~X., Woosley, S.~E., Hartmann, D.~H., \& Hoffman, R.~D. 1996, \apj, 464, 332
\bibitem[The \etal 1996]{c:the96} The, L.-S., \etal 1996, \aaps, 120, 357
\bibitem[The \etal 2006]{c:the06} The, L.-S., \etal 2006, \aap, in press, preprint in astro-ph/0601039
\bibitem[Thielemann \etal 1996]{c:thielemann96} Thielemann, F.~K., Nomoto, K., \& Hashimoto, M. 1996, \apj, 460, 408
\bibitem[Thorstensen \etal 2001]{c:thorstensen01} Thorstensen, J.~R., Fesen, R.~A., \& van den Bergh, S. 2001, \aj, 122, 297
\bibitem[Ubertini \etal 2003]{c:ubertini03} Ubertini, P., \etal 2003, \aap, 411, L131
\bibitem[Vink \etal 1996]{c:vink96} Vink, J., Kaastra, J.~S., \& Bleeker, J.~A.~M. 1996, \aap, 307, L41
\bibitem[Vink \etal 2001]{c:vink01} Vink, J., Laming, J.~M., Kaastra, J.~S., Bleeker, J.~A.~M., Bloemen, H., \& Oberlack, U. 2001, \apj, 560, L79
\bibitem[Vink \& Laming 2003]{c:vink03} Vink, J., \& Laming, J.~M. 2003, \apj, 584, 758
\bibitem[Vink 2004]{c:vink04} Vink, J. 2004, New Astronomy Review, 48, 61
\bibitem[Vink 2005]{c:vink05} Vink, J. 2005, AdSpR, 35, 976
\bibitem[Vink 2006]{c:vink06} Vink, J., \etal 2006, in preparation
\bibitem[Wietfeldt \etal 1999]{c:wietfeldt99} Wietfeldt, F.~E., Schima, F.~J., Coursey, B.~M., \& Hoppes, D.~D. 1999, \prc, 59, 528
\bibitem[Winkler \etal 2003]{c:winkler03} Winkler, C., \etal 2003, \aap, 411, L1
\bibitem[Woosley \etal 1995a]{c:wlw95a} Woosley, S.~E., Langer, N., \& Weaver, T.~A. 1995a, \apj, 448, 315
\bibitem[Woosley \& Weaver 1995b]{c:ww95b} Woosley, S.~E., \& Weaver, T.~A. 1995b, \apjs, 235, 101
\bibitem[Young \etal 2006]{c:young06} Young, P.~A., \etal 2006, \apj, 640, 891

\end{thebibliography}
\end{document}